\documentclass{aa}  

\usepackage{graphicx}

\usepackage{natbib, twoopt}
\usepackage[varg]{txfonts}

\bibpunct{(}{)}{;}{a}{}{,} 

\def\hii{\relax \ifmmode {\mbox H\,{\scshape ii}}\else H\,{\scshape ii}\fi}

\begin{document} 

    \title{Integrated-light analyses vs. colour-magnitude diagrams - II. Leo~A, an extremely young dwarf in the Local Group}

    \author{T. Ruiz-Lara \inst{1, 2}, C. Gallart \inst{1, 2}, M. Beasley \inst{1, 2}, M. Monelli \inst{1, 2}, E. J. Bernard \inst{3}, G. Battaglia \inst{1, 2}, P. S\'anchez-Bl\'azquez \inst{4}, E. Florido \inst{5, 6}, I. P\'erez \inst{5, 6}, I. Mart\'in-Navarro \inst{7, 8}}

    \authorrunning{T. Ruiz-Lara et al.}
    \titlerunning{Leo~A, an extremely young dwarf in the Local Group}

    \institute{
      \inst{1} Instituto de Astrof\'isica de Canarias, Calle V\'ia L\'actea s/n, E-38205 La Laguna, Tenerife, Spain \\
       \email{tomasruizlara@gmail.com} \\
      \inst{2} Departamento de Astrof\'isica, Universidad de La Laguna, E-38200 La Laguna, Tenerife, Spain \\
      \inst{3} Universit\'e C\^ote d'Azur, OCA, CNRS, Lagrange, France \\
      \inst{4} Departamento de F\'isica Te\'orica, Universidad Aut\'onoma de Madrid, E-28049 Cantoblanco, Spain \\ 
      \inst{5} Departamento de F\'isica Te\'orica y del Cosmos, Universidad de Granada, Campus de Fuentenueva, E-18071 Granada, Spain \\
      \inst{6} Instituto Carlos I de F\'isica Te\'orica y computacional, Universidad de Granada, E-18071 Granada, Spain \\ 
      \inst{7} University of California Observatories, 1156 High Street, Santa Cruz, CA 95064, USA \\
      \inst{8} Max-Planck Institut f\"ur Astronomie, Konigstuhl 17, D-69117 Heidelberg, Germany \\
}

    \date{Received ---; accepted ---}

 
 \abstract
     {Most of our knowledge on the stellar component of galaxies is based on the analysis of distant systems and comes from integrated light data. It is important to test whether the results of the star formation histories (SFH) obtained with standard full-spectrum fitting methods are in agreement with those obtained through colour-magnitude diagram (CMD) fitting (usually considered the most reliable approach).}
     {We compare SFHs recovered from both techniques in Leo~A, a Local Group dwarf galaxy whose majority of stars formed during the last 8 Gyrs. This complements our previous findings in a field in the Large Magellanic Cloud bar, where star formation has been on-going since early epochs though at varying rates.}
     {We have used GTC/OSIRIS in long-slit mode to obtain a high-quality integrated light spectrum by scanning a selected region within Leo~A, for which a CMD reaching the old main mequence turn-off (oMSTO) is available from HST. We compared the SFH obtained from the two datasets, using state-of-art methods of integrated light ({\tt STECKMAP}) and resolved stellar population analysis. In the case of the CMD, we computed the SFH both from a deep CMD (observed with HST/ACS), and from a shallower one (archival data from HST/WFPC2).}
     {The agreement between the SFHs recovered from the oMSTO CMD and from full spectrum fitting is remarkable, particularly regarding the time evolution of the star formation rate. The overall extremely low metallicity of Leo~A is recovered up to the last 2 Gyrs, when some discrepancies appear. A relatively high metallicity found for the youngest stars from the integrated data is a recurring feature that might indicate that the current models or synthesis codes should be revised, but that can be significantly mitigated using a more restrictive metallicity range. We thoroughly inspect the robustness of both approaches separately, finding that the subtle differences between them are inherent to the methods themselves. The SFH recovered from the shallow CMD also presents differences with the other two.}
     {Modern full-spectral fitting codes are able to recover both average constant SFHs (LMC case) and SFHs with a dominant fraction of young stellar populations. The analysis of high S/N spectra seems to provide more reliable SFH estimates than that of CMDs not reaching the oMSTO. The comparison presented in this paper needs to be repeated for predominantly old systems, thus assessing the performance of full-spectrum fitting for a full range of SFHs.}


    \maketitle
%

\section{Introduction}
\label{sec:intro}

Stars are one of the main constituents of the baryonic component of galaxies and, as a consequence, the 
characterisation of how the rate of star formation and the stellar chemical 
composition vary as a function of time is key to understanding their evolution. 
The determination of these so called Star Formation Histories (SFH) of galaxies 
\citep[e.g.][]{1973ApJ...179..427S, 1984ApJ...284..544G, 1989ARA&A..27..139H, 1996MNRAS.283.1388M, 2003MNRAS.341...33K, 2009ARA&A..47..371T, 2015ApJ...811L..18G} can 
help us to constrain their cosmological assembly history and trace back 
past events in their evolution. However, the derivation of SFHs based on the 
distribution of stars in a deep Colour-Magnitude 
Diagram (CMD) reaching the oldest main sequence turn-off \citep[oMSTO,][]{1992ApJ...388..400B, 1999AJ....118.2245G, 1999MNRAS.304..705H, 1999AJ....118.2262H, 1999AJ....117.2244O, 2002MNRAS.332...91D, 2002AJ....123.3154D, 2007ApJ...659L..17C, 2009AJ....138..558A, 2010AdAst2010E...3C}, often considered the more reliable approach, is only applicable to 
the few dozens of nearby systems found in and around the Local Group \citep[within distances up to $\sim$ 1 Mpc,][]{2012AJ....144....4M}. Thus, the study 
of the stellar content of most galaxies relies on the information that 
we can obtain from integrated light \citep[][]{2004ApJS..152..175M, 2007MNRAS.375L..16C, 2010ApJ...709...88O, 2011A&A...529A..64P, 2011MNRAS.415..709S, 2013ApJ...764L...1P, 2013A&A...557A..86C, 2014ApJ...788...72G, 2014A&A...570A...6S, 2015MNRAS.451.3400B, 2015A&A...581A.103G, 2016MNRAS.456L..35R, 
2017MNRAS.468.1902Z}.

In the last few decades, there has been  an enormous effort to improve the 
recovery of reliable SFHs from integrated spectroscopic data. Advances in the 
modeling of stellar populations \citep[e.g.][]{2003MNRAS.344.1000B, 
2005ApJS..160..176L, 2007ApJS..171..146S, 2009ApJ...699..486C, 2010MNRAS.404.1639V, 
2016MNRAS.463.3409V} based on the improvement of stellar libraries 
\citep[e.g.][]{2001A&A...369.1048P, 2003A&A...402..433L, 2004ApJS..152..251V, 
2006MNRAS.371..703S, 2007astro.ph..3658P}, isochrones and evolutionary tracks 
\citep[e.g.][]{2000A&AS..141..371G, 2004ApJ...612..168P, 2012MNRAS.427..127B, 
2013A&A...558A..46P}; studies to properly characterise the shape of the stellar 
Initial Mass Function \citep[IMF; ][]{1955ApJ...121..161S, 1996ApJS..106..307V, 
2001MNRAS.322..231K, 2013MNRAS.436.3309W, 2013MNRAS.435.2274W, 2014ApJ...784..162P} 
and the effect that these shapes have on the observed stellar populations 
\citep[][]{2015MNRAS.447.1033M}; and the development of new inversion codes to recover the star formation history from the integrated spectra of galaxies \citep[][]{2005MNRAS.358..363C, 2006MNRAS.365...46O, 2006MNRAS.365...74O, 
2009A&A...501.1269K, 2016RMxAA..52...21S, 2016RMxAA..52..171S} have substantially increased our capability to obtain the SFHs 
of external galaxies. However, astronomers are still assessing to what extent we can 
rely on the outcome of the analysis of integrated spectroscopic data.

A good number of works have tested the consistency between the SFHs, average ages 
and metallicities derived using integrated information and those using 
other methods such as the analysis of CMDs. However, most of these studies 
were focused on single stellar populations such as stellar clusters 
\citep[e.g.][]{1999AJ....118.1268G, 2002MNRAS.336..168B, 2006MNRAS.366..295D, 
2006A&A...448.1023S, 2007ApJ...655..179W, 2007MNRAS.379.1618M, 2010MNRAS.403..797G, 2010ApJ...709...88O, 
2014MNRAS.440.2953B, 2016A&A...593A..78K}, and just a few analysed more 
complex systems such as dwarf galaxies \citep[][]{2010MNRAS.406.1152M, 
2012MNRAS.423..406G} with available (albeit shallow) CMDs. Nevertheless, to fully 
assess the reliability of the information recovered from integrated light data, it 
is crucial to check this consistency analysing the most complex systems where this 
comparison can be properly done, and for which CMDs reaching the oMSTO can be obtained, 
covering a wide range of SFHs and metal enrichments to test the limitations of the methods.

In \citet[][hereafter Paper I]{2015A&A...583A..60R}, we started a project aimed at testing the 
performance of some of the most used full-spectrum fitting codes at recovering 
complex SFHs. As a case example, we derived the SFH of a region within the bar of the 
Large Magellanic Cloud (LMC) for which a CMD reaching the oMSTO and high quality 
spectroscopic data was available \citep[see also][for preliminary studies using the 
same dataset]{2002Ap&SS.281..109A, 2005astro.ph..7303L}.  
Previous studies have found different star forming episodes since the formation of the LMC bar and until the present time, with some periods of low star formation at intermediate ages \citep[][]{1999AJ....118.2262H, 1999AJ....117.2244O, 2002ApJ...566..239S, 2013MNRAS.431..364W, 2018MNRAS.473L..16M}, accompanied by a continuous chemical enrichment especially concentrated at early times and during the last few gigayears \citep[][]{2008AJ....135..836C, 2009AJ....138.1243H, 2013MNRAS.431..364W, 2018MNRAS.473L..16M}. We compared the SFHs recovered from 
the CMD using the IAC-star/MinnIAC/IAC-pop set of routines 
\citep[][]{2004AJ....128.1465A, 2009AJ....138..558A, 2011ApJ...730...14H, 
2010ApJ...720.1225M} with those recovered from the spectrum applying three modern 
full-spectrum fitting codes, namely {\tt STECKMAP} 
\citep[][]{2006MNRAS.365...74O,2006MNRAS.365...46O}, {\tt STARLIGHT} 
\citep[][]{2005MNRAS.358..363C}, and {\tt ULySS} \citep[][]{2009A&A...501.1269K}. 
{\tt STECKMAP} gave the best agreement with the CMD results. The only way of 
obtaining comparable results using {\tt STARLIGHT} or {\tt ULySS} was by using 
complex stellar populations (considering continuous star formation over an extended period of time), instead of simple ones,  as input stellar models. This last approach has subsequently been adopted in some 
works \citep[e.g.][]{2017A&A...607A.128G}. Although the agreement between SFHs 
derived from the CMD and from the spectrum was reassuring, the SFH of the LMC bar is 
just one case among the wide variety of behaviors exhibited by galactic systems. The 
natural continuation of such work is the analysis of other stellar systems sampling a wide range of evolutionary histories.

One of the most striking dwarf galaxies in the Local Group is Leo~A, a dwarf 
irregular for which recent works suggest that could have been almost purely gaseous 
during the epoch of giant galaxy assembly \citep[z~$\sim$~2,][]{2007ApJ...659L..17C}. Leo~A is 
an isolated galaxy with a well-reported population of young and massive stars 
\citep[][]{1984AJ.....89.1160D, 1996ApJ...462..684T}. Using an HST CMD 
reaching the oMSTO, \citet[][]{2007ApJ...659L..17C}  
\citep[see also ][]{2015ApJ...811L..18G} found that around 80\% of the star formation in 
Leo~A occurred within the last 8 Gyr of evolution, with a peak on the Star Formation 
Rate (SFR) between 1 to 3 Gyr  ago and an almost constant low metallicity 
(Z\footnote{Throughout the paper we use {\it Z} to denote the metallicity of the stellar component of Leo~A (unless expressed otherwise). The following equations can be used to transform from {\it Z} to [M/H] or [Fe/H] \citep[][]{2015MNRAS.449.1177V}: \begin{equation} \rm [M/H] = [Fe/H] + A\times[Mg/Fe]; \hspace{0.2cm} A \sim 0.75 \end{equation} \begin{equation} \rm [M/H] \simeq log_{\rm 10}(Z/Z_\odot) \end{equation}}~$\sim$~0.0008$\substack{+0.0005 \\ -0.0003}$). The discovery of RR Lyrae variables 
in Leo~A support the presence of a modest amount of old stars (older than 10 Gyr) coexisting with the 
dominant young component \citep[][]{2002AJ....123.3154D}. The characteristics of 
the Leo~A SFH, in comparison with the more extended SFH of the LMC bar (analysed in Paper I), make of this galaxy a key example to continue testing the 
performance of full-spectrum fitting techniques in systems with different stellar 
compositions. The results of this comparison will allow us to check, in particular, 
the ability of the integrated techniques to single out different fractions of old 
population in systems with abundant young and intermediate-age population.

In this work we obtain and compare the SFHs computed using deep CMDs and GTC/OSIRIS long slit spectroscopic data of a region within Leo~A as an example of a predominantly young galactic system. Section~\ref{sec:obs_general} describes the observations and the data reduction of the spectroscopic and photometric data used in this study. The determination of the SFH from those datasets is presented in Sect.~\ref{sec:sfh_general}. The discussion and main conclusions are outlined in Sect.~\ref{sec:discussion} and~\ref{sec:conclusions}.

\section{Observations and data reduction}
\label{sec:obs_general}

\begin{figure*}
\centering\includegraphics[width=0.55\textwidth]{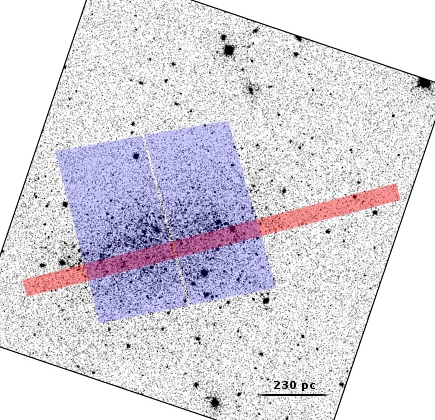} \\
\caption{Observational layout: SDSS $r$-band image \citep[][]{, 2009ApJS..182..543A} of Leo~A with the fields analysed in this works superimposed. The blue shaded area corresponds to the HST/ACS field. The red shaded area represents the scanned region observed with GTC/OSIRIS. The area of overlap between the two datasets is the comparison area from where both SFHs are extracted.}
\label{fig:obs_layout}
\end{figure*}

In this project we compare the SFH of Leo~A derived with two of the main approaches to study the stellar content of galaxies: full-spectrum fitting techniques applied to high-quality spectra and the modelling of CMDs reaching the oMSTO. Leo~A is one of the few close dwarf galaxies for which old main sequence stars can be resolved via HST photometry allowing for the observation of a deep CMD. At the same time, despite Leo~A's low surface brightness, the light collecting power of the new generation of giant telescopes makes possible to obtain a high-quality integrated spectrum. To derive the SFH following both approaches we carefully selected an extended region in this galaxy for which HST/ACS data are available, avoiding as much background and foreground objects as possible. We scanned the selected region using GTC/OSIRIS in its long slit configuration (see Fig.~\ref{fig:obs_layout}) to obtain the integrated spectrum. In this section we give all the details concerning the analysed data.

\subsection{Integrated light spectrum}
\label{sec:obs_spec}

The region within Leo~A selected to obtain the integrated spectrum was observed using the OSIRIS imager and spectrograph\footnote{Complete information regarding the OSIRIS instrument can be found at \url{http://www.gtc.iac.es/instruments/osiris/}} mounted at the Gran Telescopio Canarias (GTC) in the Observatorio del Roque de los Muchachos, La Palma. The combination of the OSIRIS instrument in its long slit configuration and the light collecting capability of a 10m-class telescope such as GTC allows for
the acquisition of high-quality spectra for objects of low surface brightness such as Leo~A \citep[$\mu_V \sim$ 24.8 mag/arcsec$^2$, averaged over one effective radius,][]{2012AJ....144....4M}. The observations were carried out as part of the \verb|GTC94-15B_000| program on December 2015, January 2016, and February 2016 using the R1000B grism and a slit width of 1.2'' that allows for a nominal wavelength coverage from 3630 to 7500~\AA~and spectral resolution of~$\sim$~11~\AA~(FWHM). This particular configuration was chosen as a compromise between the wavelength range covering the blue part, the amount of light gathered from the faint target (slit width), and the spectral resolution. However, due to peculiarities in this particular program the definitive useful wavelength range was restricted from 4000 to 5000~\AA~(see below).

The final scanned region was determined by 16 different slit positions located side by side on the sky, aligned with the semimajor axis of Leo~A (position angle of -76.1$^{\rm o}$) covering a total area of 19.2''$\times\sim$ 7'. Two separate exposures of 1800 seconds each were taken at each slit position (16 hours on target). The size of the Leo~A scanned field has been determined taking into account some crucial considerations: i) the total area has to contain enough stars to reliably determine the SFH from the CMD and to avoid under-sampling of some minority stellar populations in the field; ii) the integration area contains enough light as to obtain a high-quality integrated spectrum by summing all spectra within Leo~A (S/N/pixel $\sim$ 60 in this case), and iii) to minimise foreground and background objects not belonging to Leo~A. An inaccurate sampling of the stellar content (such as including few young stars dominating the final spectrum but with little mass contribution) may bias the reconstructed SFH. We analysed this problem in Paper I, and concluded that sampling is not a problem when integrating the light from fields that contain enough stars to reliably determine the SFH from CMD data. \citet[][]{2011ApJ...730...14H} concluded that $\simeq$~15000~stars down to the oMSTO in the CMD of a stellar system should be sufficient to accurately determine its SFH. For a system with a considerable young population as Leo~A, it is essential to have such a large number of stars in the CMD in order to have a reliable determination of the young SFH. With the 16 observed slit positions, we could analyse an area in common with the HST/ACS data of $\simeq$19.2''$\times$202'' (the second number being the width of the ACS field of view) which contains 16253 stars in the CMD (see Sect.~\ref{sec:obs_cmd}).

Classical reduction procedures such as bias subtraction, flat-fielding, wavelength calibration, cosmic rays removal \citep[L.A. Cosmic,][]{2001PASP..113.1420V} and sky subtraction were performed separately in each of the two CCDs composing the OSIRIS field of view using an IDL/python-based reduction pipeline designed to reduce OSIRIS long slit spectroscopic data. In particular, the sky subtraction is an essential step when dealing with low-surface brightness targets and deserves special consideration. 

The typical sky brightness in La Palma is $\rm \mu_V\sim$ 21.9 $\rm mag/arcsec^{2}$ \citep[][]{1998NewAR..42..503B}, i.e. three magnitudes brighter than Leo~A ($\rm \mu_V \sim$ 24.8 $\rm mag/arcsec^{2}$). In addition, the relatively low spectral resolution of our data further complicates the proper recovery of the shape of the sky lines. As a consequence, we have decided to apply a method that optimizes the usage of the data, the {\it Kelson's sky subtraction algorithm} \citep[][]{2003PASP..115..688K}. This technique relies on the knowledge of the CCD distortions and the curvature of the spectral features to obtain a characteristic sky spectrum from carefully selected pixels with sky information (avoiding bright objects as well as Leo~A contamination). We noted that some pixels close to the edge of the CCDs were affected by some illumination issue that was also visible in the exposures of the spectrophotometric standards. We corrected this issue by fitting a smooth, low-order polynomial as a function of pixel position in order to raise the affected sky pixels to the ``real'' sky level. After that, a light profile along the slit clearly showed the extension of Leo~A, its light distribution, and two nearly flat regions at both sides of the object. We applied the Kelson's algorithm to a subset of the pixels in those sky regions (one for each part of the galaxy) to properly subtract the sky on our target.

We performed several sky subtraction tests modifying the input parameters for the sky spectrum computation as well as changing the sky regions to test possible effects of this choice. We also tried different sky subtraction approaches. While most of the features in the blue part of the spectrum remain largely unaltered, the sky subtraction towards the longest wavelengths (5500~\AA~redwards) was very unstable. In addition, some clear residuals were left in the red end of the spectrum as a consequence of the large amount of sky spectral features and sky flux. For this reason, and in order to avoid the effect of an inaccurate sky subtraction, we have decided to restrict our analysis to wavelengths bluer than 5500~\AA.

\begin{figure}
\centering\includegraphics[width=0.49\textwidth]{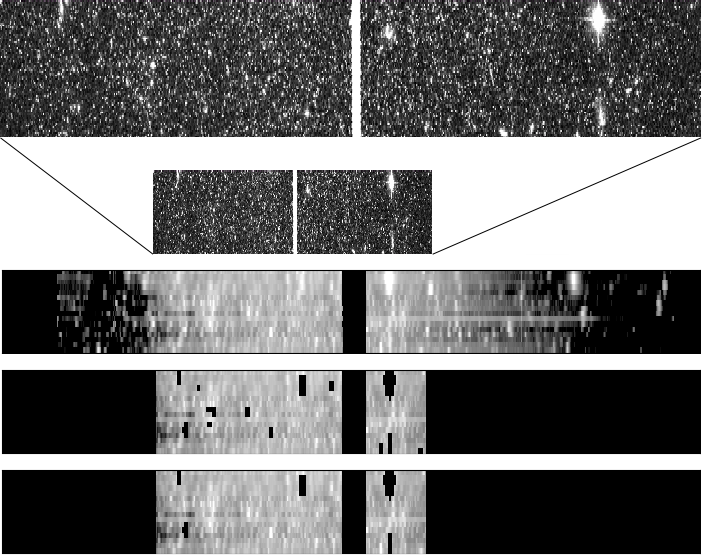} \\
\caption{Analysed region and masking procedure. Top two rows: High-spatial resolution HST image of the scanned region within the ACS (area covered is $\simeq$19.2''$\times$202''). The next three rows show pseudo-images reconstructed from the 16 slit positions observed with OSIRIS in Leo~A. Third row: The entire extension of the observed field ($\simeq$19.2''$\times$7'). Fourth row: The region from which the integrated spectrum is obtained; in this case we mask background galaxies, saturated stars, and Milky Way stars candidates (extraction A). Fifth row: The region from which the integrated spectrum is obtained; in this case we mask just foreground galaxies and saturated stars (extraction B).}
\label{fig:pseudo_masking}
\end{figure}

Figure~\ref{fig:pseudo_masking} shows a comparison of the scanned region as seen with the HST (top two rows) and a reconstruction from the 16 slit positions observed with OSIRIS (third row). Despite the difference in spatial resolution some patterns can be identified in both images. The final integrated spectrum is obtained as a mean of all the pixels within the ACS field of view avoiding saturated stars, background galaxies, and bright stars that might belong to our Milky Way rather than to Leo~A (see Sect.~\ref{sec:obs_cmd}). We will name this spectral extraction, extraction A. For this first extraction, the Milky Way candidates are selected based on their position on the observed CMD as bright stars suspected of not belonging to Leo~A (see Sect.~\ref{sec:sfh_cmd}). The low spectral resolution of the OSIRIS data hampers a proper characterisation of the membership of those bright stars using their radial velocities. As a consequence, we decided to perform a second extraction including those Milky Way candidates to further investigate the effect of these bright stars in the final SFH reconstruction. The effect of including or not these few stars is negligible in the CMD analysis, but their light could potentially affect the observed integrated spectrum (see fifth row, extraction B). We must bear in mind that this Milky Way contamination is an issue naturally found in this test due to the large area covered, but negligible in the analysis of external galaxies, typically subtending a much smaller area in the sky, where the tested codes are meant to be used.

\begin{figure}
\centering\includegraphics[width=0.49\textwidth]{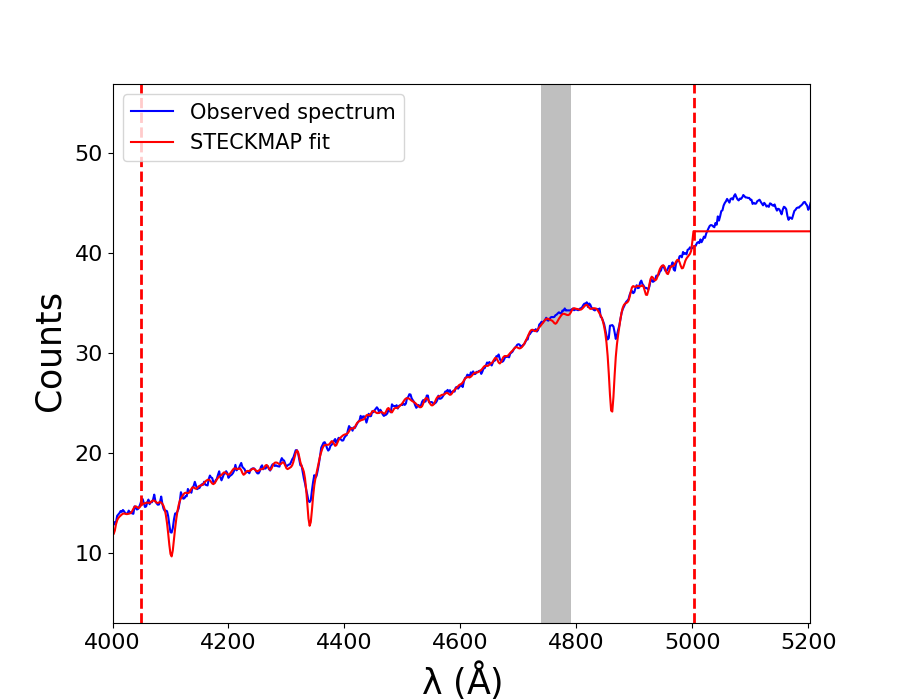} \\
\caption{Example of a typical {\tt STECKMAP} fit to the Leo~A integrated spectrum. We represent the observed spectrum with a blue solid line, the best {\tt STECKMAP} fit with a red solid line, the limits of the fit with vertical red-dashed lines, and the masked region with shade.}
\label{fig:steckmap_fit}
\end{figure}

The final extracted spectrum is shown in Fig.~\ref{fig:steckmap_fit}. A visual inspection shows another issue related with the low surface brightness of the scanned region. At $\sim$ 4760~\AA~and 5080~\AA~the integrated spectrum (regardless of the extraction procedure) displays two ``bumps'' that, in principle, do not seem to be linked to any physical feature (of stellar, gaseous, or molecular origin). A careful inspection of the sky-subtracted frames (for all slit positions) suggests that every pixel belonging to extremely to lowest surface brightness regions of the galaxy display light enhancements around such wavelengths for all the slit positions. In addition, we found this issue even in the spectrophotometric standard frames, in those pixels exposed to extremely low amounts of light (especially in the so called first CCD, CCD1). However, there is no hint of such light increases in regions where stars or background galaxies are located, i.e. in those pixels with high signal. Since these ``bumps'' are not physically connected with the Leo~A content, we have decided to avoid those wavelengths affected by these unreal features for the SFH recovery. In addition, the extremely low surface brightness of the scanned region along with the low sensitivity of the OSIRIS instrument using the R1000B grism below 4000~\AA~further restrict the spectral range useful for our purposes. 

Despite all the described difficulties, the observed spectrum is of exceptional quality from $\sim$~4000~\AA~to~$\sim$~5000~\AA, with the exception of a region around the ``bump'' prior to H$\rm \beta$. Although a better spectral sampling would be ideal for the SFH recovery from spectroscopic data, this final useful range contains some of the most important Hydrogen Balmer lines (H$\rm \beta$, H$\rm \gamma$, and H$\rm \delta$, prominent absorption lines in the figure), as well as some iron and calcium absorption features. Taking into account the goal of this work, the limited wavelength range under analysis can be interpreted as an extra challenge for testing modern full-spectrum fitting techniques rather than a caveat of this analysis. The final S/N of the integrated spectrum is around 60 (per pixel, $\sim$ 100 per \AA) with a measured resolution of around 10.8~\AA~(FWHM). A visual inspection of the observed spectrum suggest the presence of a young stellar population in Leo~A (deep Balmer lines) with some emission that can be clearly seen filling in the three Balmer lines covered by our data.

\subsection{Resolved stellar populations}
\label{sec:obs_cmd}

The CMD analysed in this work was obtained from Leo~A HST/ACS observations collected from 26$^{th}$ December 2005 to 8$^{th}$ January 2006. These data were used in \citet[][]{2007ApJ...659L..17C} to obtain for the first time the Leo~A SFH from a deep CMD reaching the oMSTO, and later on to search for variable stars \citep[][]{2013MNRAS.432.3047B}. Sixteen HST orbits were devoted to obtain precise photometry in two different bands (F475W and F814W). In each of the 16 orbits two exposures of $\sim$1200s were taken per band, accounting for a total of 19200 and 19520 seconds of integration time in the F475W and F814W bands, respectively.

The photometry of the individual stars was obtained with the {\tt DAOPHOT/ALLFRAME} set of routines \citep[][]{1994PASP..106..250S} to the non-drizzled HST/ACS images. The completeness and photometry errors were characterized through artificial star tests. A total of $\sim$ 8$\times$10$^5$ stars were used \citep[see][for details on the procedure used to obtain the photometry and the artificial stars test]{2010ApJ...720.1225M}.

\begin{figure}
\centering\includegraphics[width=0.49\textwidth]{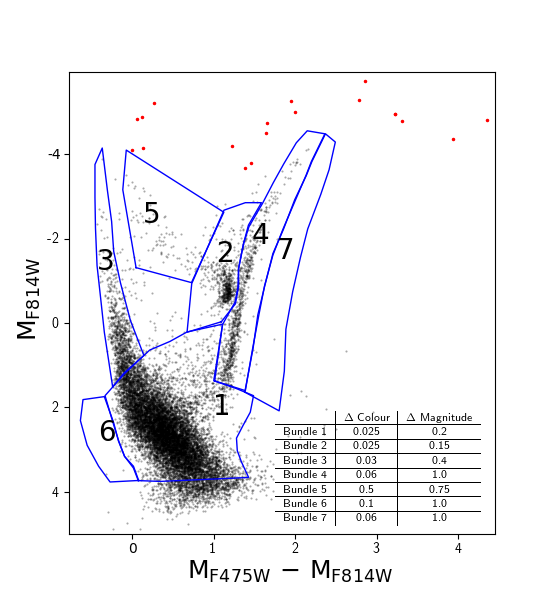} \\
\caption{(M$_{\rm F814W}$ , M$_{\rm  F475W}$ - M$_{\rm  F814W}$) CMD based on the ACS data. Red points depict the Milky Way star candidates that differentiate extraction A from extraction B (see Sect.~\ref{sec:obs_spec}). The 7 polygons show the ``bundles''. The inset table shows the size of the boxes within ``bundles''.}
\label{fig:LeoA_CMD}
\end{figure}

Although these observations comprised photometric information for some 9.5$\times$10$^4$ stars in total, in this work we are mainly interested in the area scanned with GTC/OSIRIS. The (M$_{\rm F814W}$ , M$_{\rm  F475W}$~-~M$_{\rm  F814W}$) CMD of the scanned region within Leo~A (see Fig.~\ref{fig:obs_layout}) is presented in Fig.~\ref{fig:LeoA_CMD}. In total, it comprises 16253 stars with precise and accurate photometric magnitude measurements. We transform from apparent to absolute magnitudes each observed star assuming a distance modulus of 24.48 mag \citep[determined in][using RRLyrae stars]{2013MNRAS.432.3047B} and a galactic extinction of A$_{\rm F475W}$ = 0.068 mag and A$_{\rm F814W}$~=~0.032~mag \citep[][]{2011ApJ...737..103S}. The photometry reaches down to 3.9 and 3.5 absolute magnitude (F475W and F814W, respectively), which corresponds to an apparent magnitude of 28.6 and 27.9, respectively with a 50\% of completeness. A quick visual inspection of Fig.~\ref{fig:LeoA_CMD} already shows the presence of a conspicuous young stellar population. The main aspects that stand out are the prominent and bright main sequence (MS) up to M$_{\rm F814W}$~$\sim-2$, the vertically extended red clump (M$_{\rm F814W}$~$\sim-1.0$), and the well defined and narrow red giant branch (RGB), suggesting a low spread in metallicity. The oMSTO is located around M$_{\rm F814W}$~$\sim$~2.5, well above our 50\% completeness level. In addition, there are some stars that, considering their position in the CMD, are suspected of not belonging to Leo~A but to the MW (red points). Extractions A and B of the integrated spectrum include and exclude these suspected foreground stars (see Sect.~\ref{sec:obs_spec}).

\section{Determination of the Leo~A Star Formation History}
\label{sec:sfh_general} 

In this paper we expand the comparison between the SFH derived from resolved and unresolved stellar populations presented in Paper I (focused on a region in the LMC bar) by analysing data from Leo~A, a complex system dominated by a young and intermediate-age stellar component \citep[][]{2007ApJ...659L..17C}. As in Paper I we compare the results of the CDM-fitting analysis to the results obtained from integrated spectra. In that paper we analysed different inversion codes and identified some of the main advantages and disadvantages of each one. In this paper, we focus in the performance of a single code, {\tt STECKMAP}.

\subsection{Integrated spectrum analysis}
\label{sec:sfh_spec}

The strategy to obtain the SFH from integrated spectra relies on the comparison between observed data and a set of modelled stellar spectra with some given prescriptions (IMF, SFH, etc.). The main differences among the variety of inversion codes are the way of treating the original data (e.g. polynomial fitting or not) and the way they compare observations with models (minimisation algorithms). In this work, the derivation of the SFH from the Leo~A integrated spectrum is based on the combination of three well-known analysis codes: {\tt pPXF} \citep[penalised pixel fitting code, ][]{2004PASP..116..138C, 2011MNRAS.413..813C}; {\tt GANDALF} \citep[Gas AND Absorption Line Fitting,][]{2006MNRAS.366.1151S, 2006MNRAS.369..529F}; and, {\tt STECKMAP} \citep[STEllar Content and Kinematics via Maximum A Posteriori likelihood, ][]{2006MNRAS.365...74O,2006MNRAS.365...46O}. Although {\tt STECKMAP} is the code that we use to ultimately recover the SFH for a given input spectrum, the other two codes play an essential role in this methodology.  First, {\tt pPXF} recovers the best combination of the MILES\footnote{The MILES models are publicly available at \url{http://miles.iac.es} and are based on the MILES empirical library of stellar spectra \citep[][]{2006MNRAS.371..703S, 2011A&A...532A..95F}} stellar population models \citep[][]{2016MNRAS.463.3409V} convolved with a given line-of-sight velocity distribution (LOSVD) to determine the optimal stellar velocity and velocity dispersion while masking areas affected by gaseous emission. Second, we make use of {\tt GANDALF} to include in the fit, apart from the already mentioned stellar populations models, additional Gaussians to deal with the emission lines and to take into account their possible contamination filling absorption features. Once the gaseous emission is modelled, we subtract from the observed spectrum those emission lines detected with a S/N above 3. The outcome of these two steps are the stellar kinematics and a pure absorption spectrum from our original data. It is in the third step when we are able to properly recover the stellar content shaping the observed spectrum by applying {\tt STECKMAP} to the emission-cleaned spectroscopic data. This code, based on a Bayesian minimization method, is able to recover the combination of stellar model templates that best resemble our observed spectrum (see Fig.~\ref{fig:steckmap_fit}), and thus, its SFH (SFR as a function of time and Age-Metallicity Relation, AMR). Although {\tt STECKMAP} does not rely on any {\it a priori} shape of the solution, it prefers smooth solutions over discontinuous ones (regularization). The smoothness of the final solution is determined by the user via the smoothing parameters: $\mu_{\rm x}$, $\mu_{\rm Z}$ and $\mu_{\rm v}$ for the SFR, AMR and LOSVD functions, respectively. In this final step we fix the stellar kinematics to those values computed using {\tt pPXF} to avoid the well-known degeneracy that arises when fitting simultaneously the velocity dispersion and the stellar metallicity \citep[][]{2011MNRAS.415..709S}. Errors in the SFH from the {\tt STECKMAP} analysis are computed via 25 Monte Carlo (MC) simulations. This methodology has been extensively used to recover the stellar content in external galaxies \citep[][]{2011MNRAS.415..709S, 2014A&A...570A...6S, 2015MNRAS.446.2837S, 2017MNRAS.470L.122P}. For a thorough description of the method, the codes, and the different input parameters we refer the reader to Paper I and \citet[][]{2017A&A...604A...4R}, along with the papers where each code is presented and described. 

The final shape of the SFH recovered by {\tt STECKMAP} might depend on the input parameters, i.e. the wavelength range under analysis, the model stellar templates, and the smoothing parameters. In this work, the wavelength range is limited by our spectroscopic data. Regarding the stellar templates, in Paper I we already demonstrated that the choice of models has a second order effect on the SFH recovery, affecting mainly the AMR. However, the selection of a reasonable set of smoothing parameters that improves the fit is a key test to be done before studying a new set of data with {\tt STECKMAP}. Due to the peculiarities of this particular experiment (scan of a wide region of a Local Group galaxy), another aspect to take into account is the possible foreground contamination from our own Galaxy. It might affect the shape of the observed spectrum and thus, that of the recovered SFH (extractions A and B, see Sect.~\ref{sec:obs_spec}). For the sake of clarity, we will focus on the solution from extraction A (which we consider the safest and more reasonable option), using the new set of MILES models\footnote{We do not apply any pre-selection on what sub-set of the MILES models to use in the fitting procedure as it is customary in most of the SFH derivations from spectra of integrated light, we use them all (except the oldest one, 14.0 Gyr). As a consequence, the age and metallicity limits as well as their sampling are determined by the models themselves \citep[see][]{2016MNRAS.463.3409V}. These age and metallicity values are:

[ages (Gyr)] $\times$ [Z] = [0.03, 0.04, 0.05, 0.06, 0.07, 0.08, 0.09, 0.1, 0.15, 0.20, 0.25, 0.30, 0.35, 0.40, 0.45, 0.50, 0.60, 0.70, 0.80, 0.90, 1.0, 1.25, 1.5, 1.75, 2.0, 2.25, 2.5, 2.75, 3.0, 3.25, 3.5, 3.75, 4.0, 4.5, 5.0, 5.5, 6.0, 6.5, 7.0, 7.5, 8.0, 8.5, 9.0, 9.5, 10.0000, 10.5, 11.0, 11.5, 12.0, 12.5, 13.0, 13.5] $\times$ [0.0001, 0.0003, 0.0006, 0.001, 0.002, 0.004, 0.0084, 0.010, 0.022, 0.027, 0.034, 0.047]

* {\tt STECKMAP} resamples these age bins when providing the solution by dividing logarithmically the age range.
} with the BaSTI \citep[][]{2004ApJ...612..168P} isochrones \citep[ages up to 13.5 Gyr,][]{2016MNRAS.463.3409V} to directly compare to the CMD analysis (see Sect.~\ref{sec:sfh_cmd}), and the parameters that maximise the quality of the fit ($\mu_{\rm x}$~=~0.1 and $\mu_{\rm Z}$~=~100). In Appendix~\ref{app:robust} we show that the solutions are very robust and mostly insensitive to changes on these parameters.

\begin{figure}
\centering\includegraphics[width=0.49\textwidth]{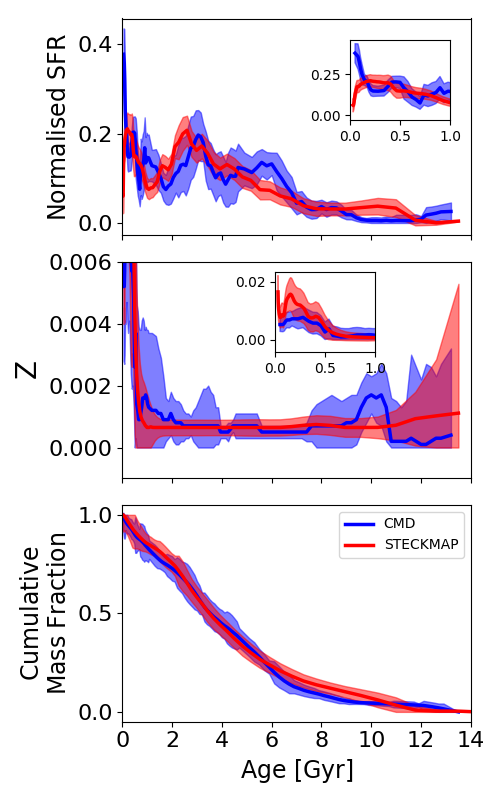} \\
\caption{Comparison between the Leo~A SFH from the CMD (blue) and the integrated spectrum using STECKMAP (red). Top panel: Normalised SFR: the area under each curve is 1. Middle panel: Age-Metallicity relation. Bottom panel: Cumulative mass fraction. See text for details on the computation of the uncertainties. Inset panels focus on the SFR and AMR at young ages.}
\label{fig:comp_1_1}
\end{figure}

Figure~\ref{fig:comp_1_1} shows the normalised SFR (top panel), the AMR (middle panel) and the cumulative mass fraction (bottom panel) from {\tt STECKMAP} (red lines and shaded areas). The normalised SFR and the cumulative mass fractions clearly show the youth of Leo~A, with only around 16\% of its mass formed more than 7 Gyr ago. Stars younger than 7 Gyr old become gradually more abundant until the SFR reaches a peak at around 2.3 Gyr. After a sharp decrement in the SFR with a minimum at $\sim$ 1 Gyr ago, the SFR increases again to reach another maximum a few hundreds of million years ago, when the SFR drops again. This SFR is basically in agreement with other works claiming a non-negligible presence of old stellar populations but with the bulk of the Leo~A population being younger than 8 Gyr \citep[][]{2002AJ....123.3154D, 2007ApJ...659L..17C, 2013MNRAS.432.3047B, 2015ApJ...811L..18G}. The AMR has a constant value (Z~=~0.0007) from 10 Gyr to 1 Gyr ago. For older populations, the recovered metallicities is around Z~=~0.0009 (with large errors, consistent with the 1-10 Gyr range), while a higher metallicity (up to Z~=~0.015) is inferred for recently formed stars. Several studies have also found such low metallicity in the stellar component of Leo~A; e.g. Z~$\sim$~0.00053 from \citet[][]{2013ApJ...779..102K}. \citet[][]{2017ApJ...834....9K} measured the metalliticy of 113 RGB stars finding metallicities between 0.0001 and 0.002, with only 2 and 6 stars with metallicities below and above this range. On the other hand, \citet[][]{2006ApJ...637..269V} measured a very low oxygen abundance in the warm gas phase in Leo~A \citep[12+log(O/H)~=~7.38~$\pm$~0.1 corresponding to 5 $\%$ of the solar oxygen content assuming 12+log(O/H)~=~~8.69~$\pm$~0.05 for the Sun;][]{2009ARA&A..47..481A}. Using this low value of the oxygen abundance as a proxy of the metallicity of the very young stars, we can conclude that the high stellar metallicity we are finding for the youngest stars using {\tt STECKMAP} seem to be in contradiction with the current gas oxygen abundance in Leo~A.

\subsection{CMD analysis}
\label{sec:sfh_cmd}

In this section, we derive the SFH of the scanned region of Leo~A through the fitting of a deep CMD obtained from the ACS data (see Sect.~\ref{sec:obs_cmd}). We use an updated python code \citep[][]{2015MNRAS.453L.113B} that closely resembles the procedures adopted in Paper I with the IAC-star/MinnIAC/IAC-pop set of routines. This method relies on the quantitative comparison, via minimization of the Poissonian equivalent to the $\chi^2$ statistics \citep[e.g.][]{2002AJ....123.3154D}, of the distribution of stars on an observed CMD with that of a combination of synthetic CMDs corresponding to simple stellar populations where observational effects have been simulated. The optimal solution is the linear combination of these simple populations best reproducing the observed CMD. We need to provide the code with a number of age and metallicity bins defining these theoretical simple populations as well as a series of regions within the CMD (``bundles'') that are used in the fitting procedure to maximise the information from the observed CMD. Each ``bundle'' is divided in boxes of different size that will determine the relative importance of each ``bundle'' in the fit. Small boxes are preferred in regions where stars in well-known evolutionary phases abound as the larger the number of boxes in a given ``bundle'', the larger the importance of it in the fit. The SFH derivation is performed in the same way as in \citet[][]{2015MNRAS.453L.113B} by counting the number of stars in each box in the observed and in the synthetic diagrams.

In this particular case, we extracted the simple stellar populations from a ``mother'', synthetic CMD \citep[][]{2009AJ....138..558A} composed of 5$\times$10$^7$ stars computed using the BaSTI stellar evolution library\footnote{We used an adapted version of the CMD code available at the BaSTI webpage (\url{http://basti.oa-teramo.inaf.it/index.html}), made available to us by S. Cassisi} \citep[][]{2004ApJ...612..168P} following a constant SFR at all ages (0.03--13.5 Gyr) and a flat metallicity distribution from Z = 0.0001 to Z = 0.01. We note here that both analyses (integrated light spectrum and CMD) use the same set of isochrones. We assume a Kroupa IMF \citep[][]{2001MNRAS.322..231K}, a binary fraction of 40$\%$ ($\beta$~=~0.4) and a minimum mass ratio q~=~0.5. We adopted this wide metallicity range rather than a more restrictive one  \citep[including just metallicities reported in the literature, e.g.][]{2007ApJ...659L..17C, 2017ApJ...834....9K} to minimise the number of assumptions that might affect the final solution. In order to take into account observational errors, this synthetic CMD was dispersed according to the artificial star tests (see Sect.~\ref{sec:obs_cmd}) and taking into account the spatial distribution of Leo~A stars. We define the different simple stellar populations according to the following age and metallicity bins, defined with similar criteria as in other SFH works using resolved CMDs \citep[][]{2010ApJ...720.1225M, 2015MNRAS.453L.113B}:

Age: [0.03, 0.1, 0.2 to 1.0 in steps of 0.2, 1.5 to 5.5 in steps of 0.5, 6.5 to 13.5 in steps of 1] Gyr

Metallicity: [1$\times$10$^{-4}$, 1.5$\times$10$^{-4}$, 2$\times$10$^{-4}$, 5$\times$10$^{-4}$, 0.001, 0.0015, 0.002, 0.004, 0.006, 0.008, 0.01] \footnote{These age and metallicity bins do not match completely the values used in the integrated spectrum analysis on purpose. We have preferred to stick to customary procedures dealing with spectrum fitting and CMD fitting techniques seeking for a proper comparison between methods.}

The adopted ``bundle'' strategy for this analysis (see Fig.~\ref{fig:LeoA_CMD}) uses the smallest boxes in the  MS and sub-giant region (``bundle'' 1), red clump (2), and bright MS (3). The RGB (4) and blue super-giant region (5) has the largest boxes along with ``bundles'' 6 and 7. Modifications in the input parameters for the fit, especially the ``bundle'' strategy or the way to deal with binaries, might affect the recovered SFH. In Appendix~\ref{app:robust} we assess how different input configurations modify the solution. Given the similarities among the SFHs derived from the different tests reported in the Appendix, we stick to the configuration that makes use of the largest number of stars from the observed CMD while still modulating the importance of different regions of the CMD in the fit by the use of the ``bundle'' strategy. 

Due to possible uncertainties in the determination of the reddening and the distance modulus as well as uncertainties in the photometric or model calibration; different solutions are computed by applying small shifts in colour and magnitude to the observed CMD. In total, a solution is calculated in each point of a grid of 25 positions within $\pm$0.06 mag and $\pm$0.15 in colour and magnitude, respectively, with a further refining of the grid around the position where the minimum $\chi^2$ is obtained. In this step each solution is computed once. After the minimum position is determined, we run 20 different solutions by i) slightly shifting the boxes and, ii) modifying the age and metallicity bins for the simple populations. The final solution is the average of these 20 solutions at the position where the minimum $\chi^2$ was found ([$\Delta$(colour)$_{\rm min}$, $\Delta$(Mag)$_{\rm min}$]). Uncertainties on the SFRs were computed as described in \citet[][]{2011ApJ...730...14H}. They are assumed to be a combination in quadrature of the uncertainties due to the effect of binning in the colour--magnitude and age--metallicity planes (from the dispersion of 20 solutions obtained after shifting the bin limits), and those due to the effect of statistical sampling in the observed CMD (the dispersion of 20 solutions obtained after resampling the observed CMD following Poissonian statistics). For a more detailed explanation of the methodology we refer the reader to \citet[][]{2010ApJ...720.1225M}, \citet[][]{2015MNRAS.453L.113B} and references therein.

The application of the above-described method to the CMD of the scanned region within the HST/ACS data for Leo~A gave a best solution (minimum $\chi^2$) that was found in [$\Delta$(colour)$_{\rm min}$, $\Delta$(Mag)$_{\rm min}$] = [0.07,-0.05] with a $\chi^2$ of 1.3. The recovered SFH is shown in Fig.~\ref{fig:comp_1_1} (blue colours). According to this solution, during the first 6 Gyr of evolution, Leo~A formed approximately 11\%~of its total mass. Afterwards, a period of star formation activity follows with three main periods of enhanced star formation with a maximum at 5.8 (lasting 3.0 Gyrs), 3.0 (extending over 1.8 Gyr), and 0.4 (active during the last 1.6 Gyr of evolution) Gyrs ago. This recovered time evolution of the SFR is in general agreement with previous works \citep[][]{2007ApJ...659L..17C, 2015ApJ...811L..18G}. Regarding the AMR, we found a rather constant and low metallicity value of 0.0005 for stars with ages from 2.5 to 10.5 Gyr with younger stars displaying higher values. As expected, given the low fraction of stars older than 10.5 Gyr, they display the larger uncertainties. On the other hand, stars younger than 2.5 present a smooth increase of their metallicities with the newly born stars having metallicities up to 0.0075. Although to a lesser degree than that found in the {\tt STECKMAP} AMR recovery, the computed metallicity for the youngest stars is above, not only previous measurements in Leo~A stars \citep[][]{2007ApJ...659L..17C, 2013ApJ...779..102K, 2017ApJ...834....9K}, but also that expected from measurement of the gaseous component \citep[][]{2006ApJ...637..269V}. However, we must bear in mind at this point that we cannot directly compare this SFH with previous determinations as in this case we are just analysing a specific region of Leo~A defined by the scanned region (see Sect.~\ref{sec:obs_spec}). 

\section{Discussion}
\label{sec:discussion}

In this section we assess the reliability of SFHs recovered via the analysis of high S/N spectra. First of all, we compare the outcome of the analysis of the Leo~A integrated spectrum and deep CMD (Sects.~\ref{sec:obs_general} and~\ref{sec:sfh_general}). Afterwards, we test the accuracy on the SFH recovery from the shallow CMD as compared to that from integrated spectra.

\subsection{Comparing integrated spectrum and CMD analysis}
\label{comparison}

Figure~\ref{fig:comp_1_1} shows the recovered SFHs after applying typical techniques to analyse integrated spectra and deep CMDs. The similarities among both recovered SFHs are remarkable, both in terms of the overall behaviour of the SFR as a function of time and the shape of the AMR. The SFR is very low from the oldest ages up to around 7 Gyr ago, when it rose until $\sim$ 3 Gyr ago. A peak of star formation is found from both approaches at $\sim$ 2.8 Gyr ago. After this peak, the SFR decreases till a minimum value around 1 Gyr ago, followed by a final increase up to the present time. Although the recovered SFH (top panel of Fig.~\ref{fig:comp_1_1}) in both analyses might differ in the details, the overall trend is remarkably similar (see bottom panel, cumulative mass fraction). It is in the AMR where some discrepancies are found. The recovered AMR is fairly constant at a value of around 0.0006 (CMD analysis) and 0.0007 (spectral analysis). Slightly higher metallicity values are inferred for young (younger than 2.5 Gyr) and old (older than 10.5 Gyr) stars than for the intermediate-age stellar populations. The behaviour at the oldest end is affected by larger uncertainties, as can be expected since very few stars/light from an old stellar population is available to constrain the solution. However, while the AMR recovered through CMD fitting increases from Z~=~0.0006, 2.5 Gyr ago, to Z~=~0.006, present day, the behaviour from the study of the integrated spectra is more abrupt, with the enrichment starting 0.5 Gyr ago and reaching up to Z~=~0.015 with some wiggles at the youngest ages. Although these discrepancies are observed and should be noted, the derived metallicities are consistent within the errors of each method.

In Paper I we already reported that the results from {\tt STECKMAP} and CMD fitting were more in agreement when considering the SFR than the AMR. In that case, the metallicities recovered from the spectral analysis were generally higher than those from the CMD analysis, with the largest discrepancies found at the youngest ages (younger than 1 Gyr), where {\tt STECKMAP} needed a large enrichment in order to properly fit the observed spectrum. Then, we speculated with the possibility that this enrichment was an artifact due to the low SFR at young ages (lack of young stars in the analysed region) or the effect of the smoothing penalty function in the AMR (large value of $\mu_{\rm Z}$). In Leo~A, where a prominent young component dominates and this issue is found regardless of the smoothing parameters (see Appendix~\ref{app:robust}), this is not a likely explanation. This issue might be related with the analysis of the integrated light or the intrinsic difficulty of current stellar models to discriminate different metallicities at young ages. 

The amount of metals in a stellar system is imprinted in its integrated spectrum on the shape of characteristic absorption spectral features. However, metal-poor and young stars tend to be hotter, which generally weakens the strength of the observed metal-lines (causing the disappearance of some). This can lead to the observation of spectra characterised by strong Balmer lines and blue, featureless continuum, not easily attributable either to a young age or a low metallicity (i.e. age-metallicity degeneracy). As a consequence, the recovery of the correct metallicity, especially for young ages, is affected by larger uncertainties. 

Apart from the intrinsic modeling issues, another possible cause for the observed differences in the AMR at young ages is the different metallicity ranges of the models employed in both approaches. While in the case of the CMD we restricted the models to vary between metallicity values of 0.0001 and 0.01 (for reasons previously explained), in the integrated light analysis we use the entire extend of the MILES models (0.0001 to 0.047). We proceed this way in order to apply both methods as independiently as possible and following the customary procedures. In the case of the analysis of the CMD, constraints on the metallicities of the system under analysis can be established either from the chemistry of individual stars \citep[][]{2017ApJ...834....9K} or by overplotting to the observed CMD isochrones of different ages and metallicities. However, nothing of that sort can be done in the case of the spectral analysis of external galaxies. In order to investigate the effect that a more restrictive metallicity range might have on the recovered solution, we have computed a different solution using only those MILES models with metallicities between 0.0001 and 0.01. Similar conclusions can be outlined regarding the shape of the temporal evolution of the SFR as before. However, in the case of the recovered AMR we can conclude that, although the maximum metallicity is now consistent with the CMD analysis (0.008), the shape is still different, implying that the chosen metallicity range cannot completely solve the problem, although it can alleviate it considerably.

If we focus on the details instead of on the overall shape of the SFH, we can find some additional discrepancies. However, at such levels of detail (of the order of time-scales of 1 Gyr), we are dealing with uncertainties within the solutions themselves (see Appendix~\ref{app:robust} for an assessment on the robustness and reliability of the two derived SFHs individually).

The main goal of this project (started in Paper I with the analysis of a field in the LMC bar) is the examination of the performance of spectral analysis at recovering the stellar content of complex systems. Leo~A is the second of such systems in which the recovered SFHs from the analysis of an integrated light spectrum and the corresponding CMD agree. As a consequence, we can conclude that both approaches produce similar solutions in systems with predominantly young (Leo~A) or approximately constant (LMC bar) star formation. In addition, they are able to discriminate different fractions of young and old stellar populations. However, we still need to assess the consistency in systems that can be considered predominantly old or dominated by intermediate-age populations. As a natural continuation of this project, we plan to repeat this comparison in other Local Group systems to consider the whole range of SFHs and chemical enrichment histories representative of galactic environments.

\subsection{On the reliability of SFHs determined from shallow CMDs}
\label{deep_shallow}

The information in the CMD of a resolved stellar system is generally regarded as the most direct way to obtain reliable information on its SFH, since stars in different evolutionary stages can be singled-out. This has led to important allocations of ground-based and HST time to obtain this kind of observations (e.g.~{\tt SMASH}, \citealt{2017AJ....154..199N}; {\tt LCID}, \citealt{2008ApJ...678L..21B}; {\tt ANGST}, \citealt{2009ApJS..183...67D}; {\tt PHAT}, \citealt{2010A&A...523A..31H}). However, there is some debate and research \citep[][]{2014ApJ...789..147W, 2015MNRAS.446.2789B} as to what extent the SFH of a stellar system, extended to its whole lifetime, can be reliably determined using a CMD which does not reach the oMSTO, but a shallower CMD reaching the magnitude level of the horizontal branch. This implies that the information on the old population is only provided by stars in advanced evolutionary stages such as the RGB, asymptotic giant branch, and horizontal branch/red clump. For systems with recent star formation, such diagrams provide information from MS stars younger than $\sim$ 1 Gyr only. Apart from that, stellar evolution models for stars in advanced evolutionary stages are affected, in principle, by larger uncertainties. Additionally, stars of different ages and metallicites are tighly packed in the CMD, where their positions are affected by important degeneracies (e.g. the well known age-metallicity degeneracy on the RGB).

In this work and in Paper I we have found that, for a young system such as Leo~A or the LMC bar, an integrated spectrum leads to a SFH that is in good agreement with that derived with a CMD reaching the oMSTO. In this section we will try to answer in the case of Leo~A, whether a shallow CMD  (where some stellar phases are missing) can also recover SFHs in close agreement with the one obtained from the deepest CMDs.

\begin{figure}
\centering\includegraphics[width=0.45\textwidth]{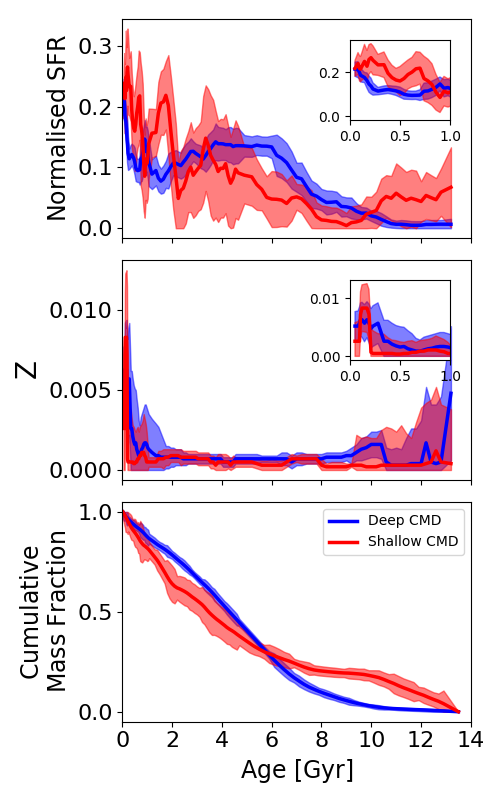} \\
\caption{Comparison of the SFHs recovered for Leo~A from the analysis of a deep CMD (blue, ACS/HST data) and a shallow one (red, WFPC2/HST data). Panels are as those in Fig.~\ref{fig:comp_1_1}. Inset panels focus on the SFR and AMR at young ages.}
\label{fig:cmd_deep_shallow}
\end{figure}

With this aim, we downloaded the photometry and artificial star tests corresponding to HST/WFPC2 observations from the HST Local Group Stellar Photometry Archive \citep[][]{2006ApJS..166..534H}, maintained by J. Holtzmann\footnote{\url{http://astronomy.nmsu.edu/holtz/archival/html/lg.html}}. In particular, we used the WFPC2 pointing covering most of the galaxy (u2x501 field, 1800 seconds in the F555W and F814W HST filters) and originally observed within the GO program 5915 (P.I. Evan Skillman). These observations are much shallower than the ones analysed previously from the HST/ACS data, reaching down to apparent magnitudes of 26.5 and 25.6 (F555W and F814W) with a 50\% of completeness. In Fig.~\ref{fig:cmd_deep_shallow} we show the recovered SFHs from both CMDs. Although the shape of the AMR is fairly well recovered (always within errors), the time evolution of the SFR displays some discrepancies, especially at old ages. Our analysis of a shallow CMD gives similar results as those published previously using similar shallow CMDs \citep[][]{1998AJ....116.1244T, 2008ApJ...686.1030O}, i.e. Leo~A is predominantly young. However, the analysis of a shallow CMD routinelly finds a non negligible amount of old population which we do not find neither with the analysis of higher-quality, deep CMDs nor from spectroscopic data (see Fig.~\ref{fig:comp_1_1}).

It is worth noting that, although integrated light analysis are also affected by the previously mentioned modeling issues, they present some advantages allowing us to understand why its analysis compares so well with the analysis of deep CMDs. Optical integrated spectra do contain light from all stellar stages (including the oMSTO) and thus, the analysis of integrated light is, in principle, sensitive to all stellar ages. However, the contribution to the integrated spectrum from stellar populations at different stages is highly dependent upon the wavelength range. At bluer wavelengths (B-band, approximately the wavelength range analysed in this work), even for old systems $\sim$ 49$\%$ of the light is coming from MS and subgiant stars with different contributions from other phases (9$\%$ Asymptotic Giant Branch, 29$\%$ RGB, 13$\%$ Horizontal Branch, according to the \citealt{2016MNRAS.463.3409V} models). As we go to redder wavelengths, RGB stars become the main contributor to the observed spectrum, losing considerably age sensitivity. This fact, added to the amount of absorption features in the blue end of the optical spectrum, makes the analysed wavelength range ideal for our purposes. Regarding the recovery of the metallicity via integrated light, we must be careful, especially in the case of the metallicity of young stars due to the age-metallicity degeneracy (as already commented before, see Sect.~\ref{comparison}).

In view of these results, we can claim that, at least for young systems such as Leo~A, the analysis of high S/N integrated spectra provides SFHs that are closer to those recovered by the analysis of a deep CMD than those from a shallow CMD. A more rigorous study analysing a larger number of systems is necessary to assess what is the optimal type of data needed to study stellar objects near the edge and beyond the LG.

\section{Conclusions}
\label{sec:conclusions}

In this paper, we apply two very distinct approaches to characterize the SFH of the central 
region of Leo~A, a dwarf irregular galaxy in the Local Group. On one hand, we scan 
the central part of Leo~A with the GTC/OSIRIS long-slit mode in order to obtain an integrated 
spectrum that has been analysed with {\tt STECKMAP}. On the other hand, we analyse 
an HST CMD reaching the oMSTO, following commonly used CMD fitting methods. 
Both approaches give remarkably consistent results in terms of the time evolution 
of the SFR in Leo~A as well as its overall chemical enrichment (AMR), If we focus on 
the details (time-scales of less than 1 Gyr), there are some differences between both 
approaches that can be attributed to uncertainties inherent to each method, and 
thus, details from both recovered SFHs have to be taken with caution. Slight differences are also found in the recovered metallicity of young stars, differences that can be highly minimised restricting the metallicity range of the used stellar models. This general 
agreement is reassuring, especially considering that most of what we currently know 
about galaxy formation and evolution comes from studies of external systems where 
individual stars cannot be resolved. Together with the conclusions in  
\citet[][]{2015A&A...583A..60R},  the results presented in this paper support the use of 
high-quality spectroscopic data collected at 
ground-based facilities to study the SFHs of external as well as nearby, semi-resolved systems for which 
a CMD reaching the oMSTO cannot be obtained. Through 
the results of the analysis of a shallow CMD (reaching just below the horizontal branch) we provide a 
hint that a high S/N spectrum like the one used in this work may be preferred to a 
shallow CMD (which are very expensive to obtain for systems beyond the LG), at least in the case of young stellar systems such as Leo~A.

\begin{acknowledgements}

We thank the anonymous referee for the careful reading of the manuscript and the very useful comments that have improved considerably this work. We also thank Alexandre Vazdekis and Jes\'us Falc\'on-Barroso for useful discussions. This research has been mainly supported by the Spanish Ministry of Economy and Competitiveness (MINECO) under the grants AYA2014-56795-P and AYA2016-77237-C3-1-P. EJB acknowledges support from the CNES postdoctoral fellowship program. GB gratefully acknowledges financial support by the Spanish Ministry of Economy and Competitiveness (MINECO) under the Ramon y Cajal Programme (RYC-2012-11537) and the grant AYA2014-56795-P. PSB thanks the support under the grant AYA2013-48226-C3-3-P (MINECO). EFN and IP thanks the support received via grants AYA 2014-53506-P (MINECO) and FQM-108 (Junta de Andaluc\'ia). IMN acknowledges funding from the Marie Sk\l odowska-Curie Individual Fellowship 702607, and from grant AYA2013-48226-C3-1-P from the Spanish Ministry of Economy and Competitiveness (MINECO). This research has been based on observations made with the Gran Telescopio Canarias (GTC), installed at the Spanish Observatorio del Roque de los Muchachos of the Instituto de Astrofísica de Canarias, in the island of La Palma as well as based on observations made with the NASA/ESA Hubble Space Telescope, obtained at the Space Telescope Science Institute, which is operated by the Association of Universities for Research in Astronomy, Inc., under NASA contract NAS 5-26555. These observations are associated with HST program \verb|10590| and GTC program \verb|GTC94-15B_000|.

This research makes use of python (\url{http://www.python.org}); Matplotlib \citep[][]{hunter2007}, a suite of open-source python modules that provide a framework for creating scientific plots; and Astropy, a community-developed core Python package for Astronomy \citep[][]{astropy2013}.

\end{acknowledgements}

\bibliographystyle{aa} 
\bibliography{bibliography} 

\appendix

\section{Robustness of the solutions}
\label{app:robust}

In the main body of this paper we have focused our comparison between the performances of full-spectral fitting and CMD-based techniques in a one-to-one comparison. This means that, based on standard criteria, we have chosen a particular set of input parameters to compute the {\it final} solution from each approach. However, the details of this solution are somewhat dependent on the set of input parameters. In this appendix we assess how different input parameters affect the recovery of the SFH. This can help us to understand the actual robustness of the method and thus, comprehend how detailed and accurate the SFH recovery actually is and to what extend we can rely on small details in the SFH shape. 

\subsection{Integrated spectra}

Table~\ref{tab:steckmap_tests} summarises the different runs performed in order to obtain the {\it final} SFH from the integrated spectrum. These runs focus on the effect of the choice of the smoothing parameters and the possible influence of bright foreground stars (Milky Way candidates). The solution that has been compared with the CMD approach was the one that minimised the residuals in the spectral fit ($rms$, run 6) from extraction A.

\begin{table*}
\centering
\begin{tabular}{ccccc}
\hline\hline
Run & $\mu_{\rm x}$ & $\mu_{\rm Z}$ & \multicolumn{2}{c}{$rms$} \\  \hline
  &   &   & Extraction A & Extraction B  \\  \hline
1  &   1     &  100     &   0.124 &     0.118  \\
2  &   10    &   10     &   0.130 &    0.119  \\
3  &   1     &    1     &   0.125 &     0.133  \\
4  &   0.1   &    0.1   &   0.129 &     0.122  \\
5  &   100   &    0.1   &   0.124 &     0.121  \\
6  &   0.1   &   100    &   0.113* &    0.121  \\
7  &   100   &   100    &   0.121 &     0.106*  \\
8  &   1     &   1000   &   0.131 &     0.120 \\
9  &   0.001 &    0.001 &   0.136 &     0.122   \\
10 &   1000  &   0.01   &   0.133 &     0.129  \\
11 &   0.01  &   1000   &   0.124 &     0.124  \\
\hline
\end{tabular}
\caption{Set of input parameters for the {\tt STECKMAP} runs to recover the SFH from the integrated spectrum. The main parameters that might affect the shape of the recovered SFH are the smoothing parameters ($\mu_x$ and $\mu_Z$) and the extraction mask (extraction A or B). With these runs we cover all reasonable values of smoothness in the solution. We show as well the $rms$, computed as the mean values of the absolute differences between the data and the fit as a proxy for the quality of each fit. The solutions with the lowest $rms$ are marked with a *.} 
\label{tab:steckmap_tests}
\end{table*}

\begin{figure}
\centering\includegraphics[width=0.49\textwidth]{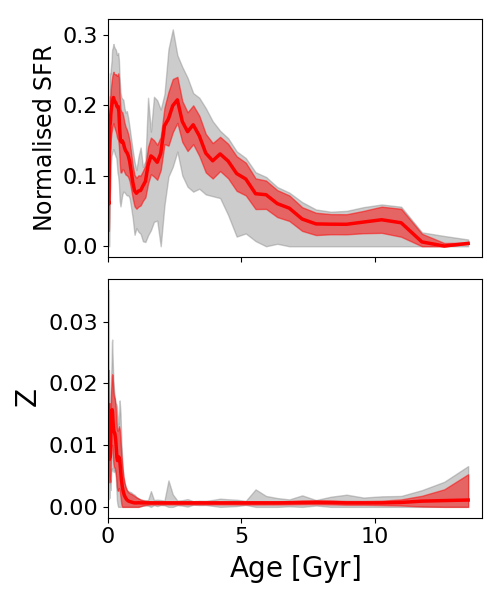} \\
\caption{Comparison of the SFHs recovered from the analysis of the Leo~A integrated spectrum using different smoothing parameters (see Table~\ref{tab:steckmap_tests}). Panels are as those in the top and middle panels of Fig.~\ref{fig:comp_1_1}. In red we show the best solution (minimum $rms$) for the extraction A (run 6). Embedded in the grey shaded area are runs 1 to 11 using extraction A, including errors.}
\label{fig:steckmap_smoothing}
\end{figure}

In order to investigate to what extent different smoothing parameters ($\mu_x$ and $\mu_Z$) affect the recovery of the SFH, we have tested 11 different sets of parameters sampling a reasonable range of values (see Table~\ref{tab:steckmap_tests}). In Fig.~\ref{fig:steckmap_smoothing} we compare the shape of the solution used throughout this paper (extraction A and minimum $rms$, red) with the envelope of the 11 solutions (including errors) using this input configuration (extraction A) but considering all the different smoothing parameter sets. We must highlight that, in spite of some differences, every recovered SFH presents a common pattern: a near absence of stars older than 7 Gyr, two peaks of stars with ages 2.3 and 0.1 Gyr, an almost flat AMR with values of Z~$\sim$~0.0007 and a chemical enrichment towards young stellar populations. This enrichment might reach up to solar values in some runs. The small differences in the $rms$ values found among runs show that the choice of parameters is not a crucial factor in this case, and proves the robustness of the {\tt STECKMAP} fits in analysing the Leo~A spectrum. Both aspects suggest that small, secondary peaks, drops and trends in the recovered SFH, apart from the above outlined general behaviour, are not sufficently robust as to be claimed as real, and might be artifacts of the particular set of input parameters.

\begin{figure}
\centering\includegraphics[width=0.49\textwidth]{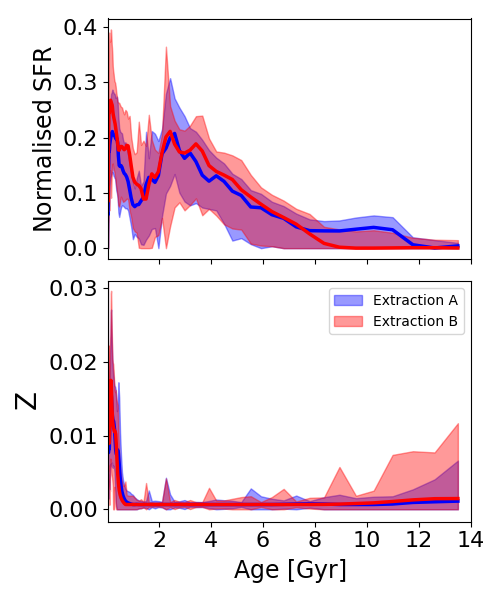} \\
\caption{Comparison of the SFHs recovered from the analysis of the Leo~A integrated spectrum using extraction A (blue) or B (red). Panels are as those in the top and middle panels of Fig.~\ref{fig:comp_1_1}. All the different solutions from the different runs, including errors (see Table~\ref{tab:steckmap_tests}), are embedded within each shaded area.}
\label{fig:steckmap_a_vs_b}
\end{figure}

For the sake of completeness, we also assess the effect that bright Milky Way candidate stars might have on the SFH recovery. Although this is not an issue affecting normal studies of stellar populations in external systems, it is an aspect that might affect the conclusions of the current work. In Sect.~\ref{sec:obs_general} we inspected the observed CMD pinpointing bright stars that might not belong to Leo~A but could have a potential effect on the shape of the observed spectrum and thus, on the recovered SFH. Figure~\ref{fig:steckmap_a_vs_b} compares the envelopes of all the solutions (including errors) for the two different extractions performed in Sect.~\ref{sec:obs_spec} (see also Fig.~\ref{fig:pseudo_masking}) as well as the best solutions in each set of runs (solid lines). Although the light coming from these bright, MW candidates slightly modifies the shape of the spectrum, the light coming from Leo~A dominates in the SFH recovery and both extractions lead to similar SFHs. The main differences are spotted at old ages (10 to 12 Gyr ago) and at very young ages. 
These minor discrepancies are somehow expected, as the position in the observed CMD of most of these stars suggests that they might be young if at Leo~A distance. As a consequence, extraction B (which includes the light from these stars) presents a slightly larger fraction of stars younger than 1 Gyr in detriment of star with ages around 10 to 12 Gyr. Despite these small differences, throughout the paper we have decided to stick our analysis to what we consider the safer and more standard extraction, extraction A.

\subsection{Resolved stellar populations}

As in the case of the SFH recovery from integrated data, there is a degree of uncertainty in the results from the analysis of the Leo~A observed CMD due to the choice of input parameters (see Table~\ref{tab:cmd_tests}). In this appendix we focus on the effect of the two main aspects that might change the recovered SFH and thus, our conclusions (see Table~\ref{tab:cmd_tests}), namely the ``bundle'' strategy (highly human-dependant) and the way of treating binaries in the computation of the synthetic CMD \citep[][]{2017MNRAS.471.2812B}. Although the purpose of this work is not a complete study of the SFH of Leo~A, in these robustness tests we have decided to analyse the entire Leo~A CMD with around 95000 stars (not only the stars within the scanned region). The larger number of stars in comparison with the scanned region allowed us to better discriminate the possible effect of the input parameters under analysis as larger errors are obtained as we reduce the number of stars. 

\begin{table*}
\centering
\begin{tabular}{cccc}
\hline\hline
Test & ``Bundle'' strategy & $\beta$ & q \\  \hline
A  &   All CMD in 7 ``bundles''     &  0.4     &   0.5  \\
B  &   A except RC and RGB    &   0.4     &   0.5  \\
C  &   All CMD in 1 single ``bundle''     &    0.4     &   0.5  \\
D  &   Same as A   &    0.7   &   0.1  \\
\hline
\end{tabular}
\caption{Set of input parameters for the tests to recover the SFH from the Leo~A observed CMD. The main parameters that might affect the shape of the recovered SFH are the ``bundle'' strategy and the way of dealing with binaries during the synthetic CMD computation. For a graphical description of the different ``bundle'' strategies see Fig.~\ref{fig:cmd_robust_bundle_strategy}.} 
\label{tab:cmd_tests}
\end{table*}

\begin{figure*}
\centering\includegraphics[width=0.95\textwidth]{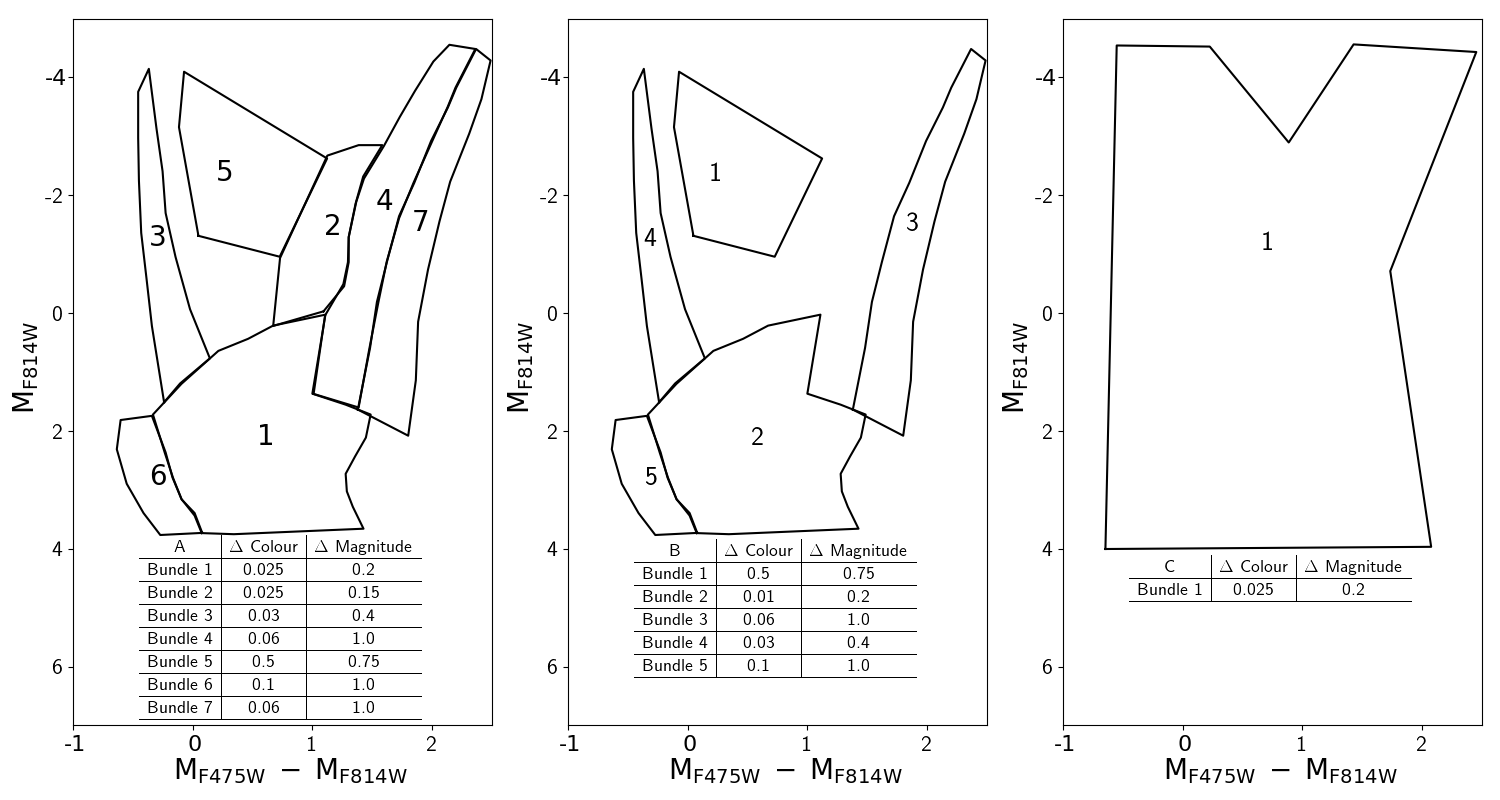} \\
\caption{Different ``bundle'' strategies adopted to test the robustness of the recovery of the SFH through CMD fitting. A vs. B allows us to tackle the effect of adding the RGB or the RC in the CMD fit. Test C allows us to check the effect of considering all the CMD at once in a single ``bundle'' with small boxes.}
\label{fig:cmd_robust_bundle_strategy}
\end{figure*}

Figure~\ref{fig:cmd_robust_bundle_strategy} shows the three different ``bundle'' strategies used in these tests. There are some areas within the CMD that are populated by stars in evolutionary phases possibly less well described by theory. As a consequence, taking into account that the recovery of the SFH relies on the comparison of observed CMDs with synthetic ones based on stellar evolution, one might think that the addition of these regions in the analysis might affect the recovered results. Test A analyses all regions with stars within the CMD in 7 different boxes including the MS, sub-giant, Red Clump (RC) and RGB stars with different boxes sizes. Test B follows the previous strategy but without the RC and RGB ``bundles''. Test C is intended as an experiment to further assess the effect of the “bundle” definition, which is rather subjective, by analysing the CMD with a single “bundle” and a fine grid containing all the stars.

\begin{figure}
\centering\includegraphics[width=0.45\textwidth]{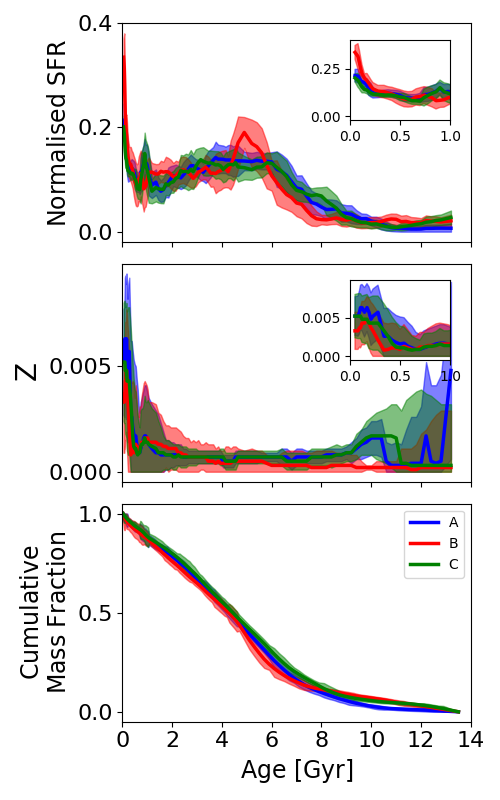} \\
\caption{Comparison of the SFHs recovered from the analysis of the Leo~A CMD using different ``bundle'' strategies (see Fig.~\ref{fig:cmd_robust_bundle_strategy}). In blue, red and green we show the SFH (and the errorrs) recovered using ``bundle'' strategies A, B and C, respectively. Panels are as those in Fig.~\ref{fig:comp_1_1}. Inset panels focus on the SFR as well as AMR at young ages.}
\label{fig:cmd_sol_robust_bundles}
\end{figure}

The outcome of all these tests are shown in Fig.~\ref{fig:cmd_sol_robust_bundles}. All solutions are consistent with a system with very low star formation until $\sim$ 8 Gyr ago, when the SFR starts to rise to a period of higher activity between 6 and 2 Gyr ago, followed by a slight decrease followed by a recent rise. The behaviour of the AMR is also very similar among tests: a flat AMR is found with the exceptions of a recent chemical enrichment and larger uncertainties at old ages, where the number of stars in which the solution is based is minimal. As the cumulative mass fraction shows, all these discrepancies are mere subtleties, and thus, the effect of the chosen ``bundle'' strategy does not affect the overall shape of the recovered SFH, but only the details. In fact, all these recovered SFHs are in agreement with previous works using the same data.

\begin{figure}
\centering\includegraphics[width=0.45\textwidth]{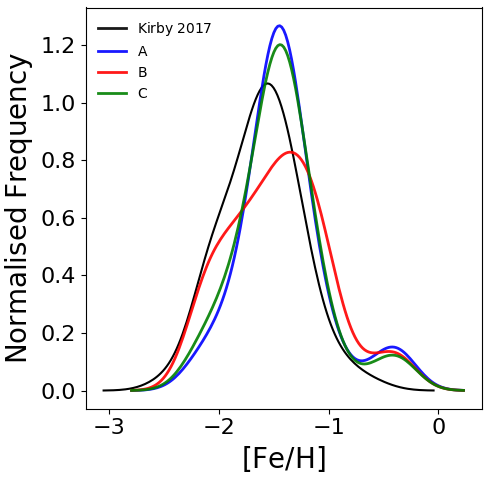} \\
\caption{Modelled and observed RGB stars metallicity distributions. We compare the metallicity distributions of the RGB stars from the best-model CMD for tests A, B and C (blue, red and green) with that presented in \citet[][]{2017ApJ...834....9K} determined through the spectroscopic analysis of individual stars (black, 113 stars). The distributions are built as the sum of individual gaussians for each star.}
\label{fig:met_distribution}
\end{figure}

The different choices of input parameters also have a small effect on the properties of the best-fit CMD. Figure~\ref{fig:met_distribution} shows the metallicity distribution of a subset of RGB stars \citep[matching the absolute magnitudes covered by][]{2017ApJ...834....9K} in the three best-fit CMDs (for tests A, B, and C; blue, red and green, respectively) in comparison with that presented in \citet[][]{2017ApJ...834....9K} from the spectroscopic chemical composition of 113 individual RGB stars (black). The resemblance between the four distributions is remarkable, although some differences can be highlighted. Tests A and C are those that present the higher similarities as compared with the \citet[][]{2017ApJ...834....9K} distribution, while test B results in a broader [Fe/H] distribution. This fact indicates that adding as many evolutionary phases as possible in our analysis improves the reliability of the recovered SFHs. In particular, if we include the observed RGB in the fit, the characteristics of the modeled RGB resembles those of the observed one (metallicity distribution) even though this stellar evolution phase is affected, in principle, by larger uncertainties. This result, in fact, may provide an indication of the reliability of current, up-to-date stellar evolution models such as those of the BaSTI library, even in advanced evolutionary stages.

\begin{figure}
\centering\includegraphics[width=0.4\textwidth]{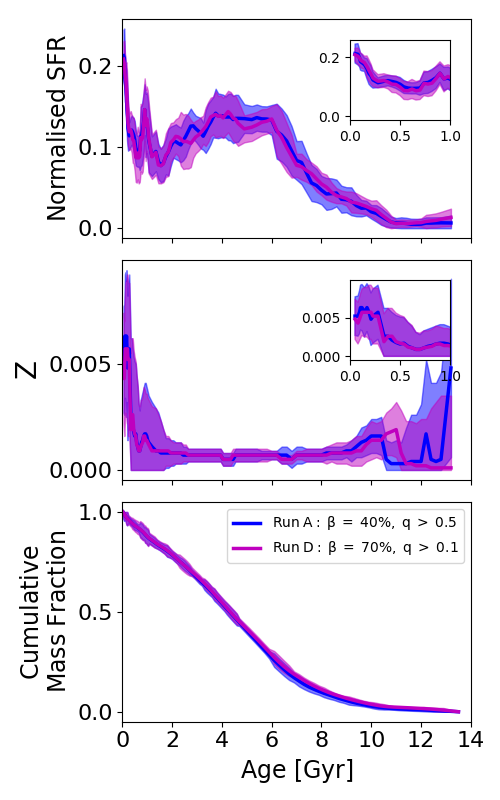} \\
\caption{Comparison of the SFHs recovered from the analysis of the Leo~A CMD with different synthetic CMDs treating binaries with different recipes (``bundle'' strategy as in test A). Panels are as those in Fig.~\ref{fig:comp_1_1}. Inset panels focus on the SFR as well as AMR at young ages.}
\label{fig:cmd_sol_robust_binaries}
\end{figure}

Another parameter that affects the distribution of stars in the CMDs is stellar multiplicity, as stars are typically born in pairs and even in groups. However, the problem of how to model the binary population is far from trivial \citep[][]{2011A&A...529A..92K, 2013pss5.book..115K, 2017MNRAS.471.2812B}. As a consequence, current codes to create synthetic CMDs are based on simple parametrisation in which properties as the binary fraction ($\beta$) or the binary mass ratio (q) are essential. Figure~\ref{fig:cmd_sol_robust_binaries} compares the SFH recovered by comparing the Leo~A observed CMD with synthetic CMDs with different $\beta$ and q values (see Table~\ref{tab:cmd_tests}). As in the previous set of tests, the recovered SFH is fully consistent in both cases. Only subtle discrepancies appear among tests and slightly younger populations are found (see bottom panel of Fig.~\ref{fig:cmd_sol_robust_binaries}) as the amount of binaries increase \citep[see also][]{2010ApJ...720.1225M}. This suggests that the overall shape of the SFH presents little dependence on this particular choice.

Based on this analysis, we conclude that all the tests produce consistent solutions from the study of the Leo~A CMD reaching the oMSTO, with no obvious ``best'' solution. Small discrepancies inherent to the method are found with different input parameters at short time-scales (of the order of 1 Gyr). In addition, this analysis opens the door of an alternative and more objective ``bundle'' definition in which just one ``bundle'' is considered.

\vspace{0.5cm}

To sum up, we can conclude that both approaches present their own typical uncertainties (usually subtle) and that the choice of input parameters within reasonable limits has a small impact. To what extent we can rely on these subtle details is an aspect to be improved within each technique in the coming years.

\end{document}